\documentclass[showpacs, nobibnotes, amsmath, prd, floatfix,
 twocolumn, superscriptaddress, apsrev4-1 ]{revtex4-1}

\usepackage{graphicx}
\usepackage{dcolumn}
\usepackage{bm}
\usepackage{color}

\newcommand{\GeVc}{GeV${}/c$}
\newcommand{\GeVcc}{GeV${}/c^2$}
\newcommand{\todo}[1]{\textcolor{red}{\textit{{#1}}}}
\newcommand{\Mgg}{M_{\gamma\gamma}}
\newcommand{\xBj}{x}  
\begin{document}

\preprint{preprint-number}

\title{Neutral pion cross section and spin asymmetries at intermediate
  pseudorapidity in polarized proton collisions at $\sqrt{s} = 200$ GeV}




\affiliation{AGH University of Science and Technology, Cracow, Poland}
\affiliation{Argonne National Laboratory, Argonne, Illinois 60439, USA}
\affiliation{University of Birmingham, Birmingham, United Kingdom}
\affiliation{Brookhaven National Laboratory, Upton, New York 11973, USA}
\affiliation{University of California, Berkeley, California 94720, USA}
\affiliation{University of California, Davis, California 95616, USA}
\affiliation{University of California, Los Angeles, California 90095, USA}
\affiliation{Universidade Estadual de Campinas, Sao Paulo, Brazil}
\affiliation{Central China Normal University (HZNU), Wuhan 430079, China}
\affiliation{University of Illinois at Chicago, Chicago, Illinois 60607, USA}
\affiliation{Cracow University of Technology, Cracow, Poland}
\affiliation{Creighton University, Omaha, Nebraska 68178, USA}
\affiliation{Czech Technical University in Prague,
 FNSPE, Prague, 115 19, Czech Republic}
\affiliation{Nuclear Physics Institute AS CR, 250 68 \v{R}e\v{z}/Prague,
 Czech Republic}
\affiliation{Frankfurt Institute for Advanced Studies FIAS, Germany}
\affiliation{Institute of Physics, Bhubaneswar 751005, India}
\affiliation{Indian Institute of Technology, Mumbai, India}
\affiliation{Indiana University, Bloomington, Indiana 47408, USA}
\affiliation{Alikhanov Institute for Theoretical and Experimental Physics,
 Moscow, Russia}
\affiliation{University of Jammu, Jammu 180001, India}
\affiliation{Joint Institute for Nuclear Research, Dubna, 141 980, Russia}
\affiliation{Kent State University, Kent, Ohio 44242, USA}
\affiliation{University of Kentucky, Lexington, Kentucky, 40506-0055, USA}
\affiliation{Korea Institute of Science and Technology Information,
 Daejeon, Korea}
\affiliation{Institute of Modern Physics, Lanzhou, China}
\affiliation{Lawrence Berkeley National Laboratory, Berkeley,
 California 94720, USA}
\affiliation{Massachusetts Institute of Technology, Cambridge,
 MA 02139-4307, USA}
\affiliation{Max-Planck-Institut f\"ur Physik, Munich, Germany}
\affiliation{Michigan State University, East Lansing, Michigan 48824, USA}
\affiliation{Moscow Engineering Physics Institute, Moscow Russia}
\affiliation{National Institute of Science Education and Research,
 Bhubaneswar 751005, India}
\affiliation{Ohio State University, Columbus, Ohio 43210, USA}
\affiliation{Old Dominion University, Norfolk, VA, 23529, USA}
\affiliation{Institute of Nuclear Physics PAN, Cracow, Poland}
\affiliation{Panjab University, Chandigarh 160014, India}
\affiliation{Pennsylvania State University, University Park,
 Pennsylvania 16802, USA}
\affiliation{Institute of High Energy Physics, Protvino, Russia}
\affiliation{Purdue University, West Lafayette, Indiana 47907, USA}
\affiliation{Pusan National University, Pusan, Republic of Korea}
\affiliation{University of Rajasthan, Jaipur 302004, India}
\affiliation{Rice University, Houston, Texas 77251, USA}
\affiliation{Universidade de Sao Paulo, Sao Paulo, Brazil}
\affiliation{University of Science \& Technology of China,
 Hefei 230026, China}
\affiliation{Shandong University, Jinan, Shandong 250100, China}
\affiliation{Shanghai Institute of Applied Physics,
 Shanghai 201800, China}
\affiliation{SUBATECH, Nantes, France}
\affiliation{Temple University, Philadelphia, Pennsylvania, 19122, USA}
\affiliation{Texas A\&M University, College Station, Texas 77843, USA}
\affiliation{University of Texas, Austin, Texas 78712, USA}
\affiliation{University of Houston, Houston, TX, 77204, USA}
\affiliation{Tsinghua University, Beijing 100084, China}
\affiliation{United States Naval Academy, Annapolis, MD 21402, USA}
\affiliation{Valparaiso University, Valparaiso, Indiana 46383, USA}
\affiliation{Variable Energy Cyclotron Centre, Kolkata 700064, India}
\affiliation{Warsaw University of Technology, Warsaw, Poland}
\affiliation{University of Washington, Seattle, Washington 98195, USA}
\affiliation{Wayne State University, Detroit, Michigan 48201, USA}
\affiliation{Yale University, New Haven, Connecticut 06520, USA}
\affiliation{University of Zagreb, Zagreb, HR-10002, Croatia}

\author{L.~Adamczyk}\affiliation{AGH University of Science and Technology,
 Cracow, Poland}
\author{J.~K.~Adkins}\affiliation{University of Kentucky,
 Lexington, Kentucky, 40506-0055, USA}
\author{G.~Agakishiev}\affiliation{Joint Institute for Nuclear Research,
 Dubna, 141 980, Russia}
\author{M.~M.~Aggarwal}\affiliation{Panjab University,
 Chandigarh 160014, India}
\author{Z.~Ahammed}\affiliation{Variable Energy Cyclotron Centre,
 Kolkata 700064, India}
\author{I.~Alekseev}\affiliation{Alikhanov Institute for Theoretical
 and Experimental Physics, Moscow, Russia}
\author{J.~Alford}\affiliation{Kent State University, Kent, Ohio 44242, USA}
\author{C.~D.~Anson}\affiliation{Ohio State University,
 Columbus, Ohio 43210, USA}
\author{A.~Aparin}\affiliation{Joint Institute for Nuclear Research,
 Dubna, 141 980, Russia}
\author{D.~Arkhipkin}\affiliation{Brookhaven National Laboratory,
 Upton, New York 11973, USA}
% ...

\collaboration{STAR Collaboration}\noaffiliation\bibliographystyle{apsrev}

\date{\today}

\begin{abstract}

The differential cross section and spin asymmetries for neutral pions
produced within the intermediate pseudorapidity range $0.8 < \eta < 2.0$ in
polarized proton-proton collisions at $\sqrt{s} = 200$ GeV are presented.
Neutral pions were detected using the endcap electromagnetic calorimeter
in the STAR detector at RHIC. The cross section was measured over a
transverse momentum range of $5 < p_T < 16$ GeV$/c$ and is found to
agree with a next-to-leading order perturbative QCD calculation. The
longitudinal double-spin asymmetry, $A_{LL}$, is measured in the
same pseudorapidity range and spans a range of Bjorken-$x$
down to $x\approx0.01$. The measured $A_{LL}$ is consistent with
model predictions for varying degrees of gluon polarization. The
parity-violating asymmetry, $A_L$, is also measured and found to be
consistent with zero. The transverse single-spin asymmetry, $A_N$, is
measured over a previously unexplored kinematic range in Feynman-$x$
and $p_T$. Such measurements may aid our understanding of the on-set
and kinematic dependence of the large asymmetries observed at more
forward pseudorapidity ($\eta\approx3$) and their underlying mechanisms.
The $A_N$ results presented are consistent with a twist-3 model prediction
of a small asymmetry over the present kinematic range.

\end{abstract}

\maketitle

\section{Introduction}

The production of $\pi^0$-mesons in $p+p$ collisions at
$\sqrt{s}=200$ GeV provides access to the combination of quark and
gluon distribution functions within the proton, coupled with the
fragmentation functions of the produced $\pi^0$. For neutral pion
production at $\sqrt{s}=200$ GeV over the intermediate
pseudorapidity range $0.8 < \eta < 2$ and the transverse momentum
range $5 < p_{T} < 16$ GeV$/c$ the quark-gluon subprocess dominates
over gluon-gluon and quark-quark subprocesses
\cite{NLOcorrectionsPion, NLOcorrectionsJet, KretzerSubprocessDep,
*ResearchPlan}. Previously published data on inclusive $\pi^0$ production
in polarized proton-proton scattering have been at either central
pseudorapidity ($-1<\eta<1$) \cite{PHENIX_pi0_xSecRun2,
*PHENIX_pi0_xSecRun3, *PHENIX_pi0_xSec200, PHENIX_pi0_ALL-03,
*PHENIX_pi0_ALL-04, *PHENIX_pi0_ALL, PHENIX_mid_AN,
STAR_pi0_BEMC, STAR_pi0_BEMC2} or at forward pseudorapidity
($\eta \approx 3 $) \cite{E704-88, *E704, E704_ALL, STAR_FPD_AN1,
STAR_FPD1_xSec, STAR_FPD2_xSec}. The measurements described
in this paper, taken at intermediate pseudorapidity, cover a less-constrained
region of the Bjorken-scaling variable, $\xBj$, and previously unmeasured
regions of the Feynman-$x$ and $p_T$ kinematic domains. Feynman-$x$
is defined as $x_{F} = 2p_{L}/\sqrt{s}$, where $p_{L}$ represents the
longitudinal momentum of the pion relative to the direction of the polarized
beam.

Global analyses of fragmentation functions have shown that, due to
increased sensitivity to gluonic scattering, RHIC measurements of
inclusive pion production at central and forward pseudorapidity have been
useful in constraining the gluon fragmentation function \cite{DSS}. Since the
present data span intermediate pseudorapidity and transverse momentum,
they are expected to be sensitive to a different mix of partonic subprocesses
than previous measurements at central and forward pseudorapidity. Thus,
comparison of the present measured cross section to perturbative QCD
(pQCD) calculations may aid current understanding of the gluon
fragmentation function. Previous cross section measurements which span a
similar range of $p_{T}$ at central pseudorapidity \cite{PHENIX_pi0_xSec200,
STAR_pi0_BEMC, PHENIX_pi0_xSec500} typically agree within the scale
uncertainty of the pQCD prediction in the region of $5 < p_T < 16$ \GeVc.

\begin{figure}
  \begin{center}
    \includegraphics[width=0.49\textwidth]{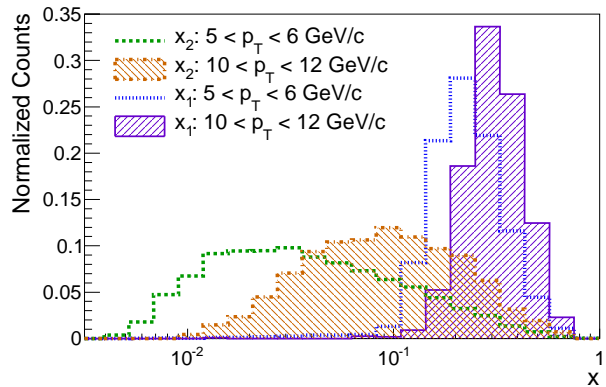}
  \end{center}
  \caption{\label{fig:x1x2}(Color online) Distributions of $x_1$ and $x_2$
  in two different bins of reconstructed $\pi^0$ $p_T$ for events at
  $\sqrt{s}=200$ GeV over $0.8<\eta<2$. The distributions were made
  using Monte Carlo simulations based on PYTHIA \cite{Pythia,Pythia2},
  utilizing unpolarized parton distribution functions.}
\end{figure}

The longitudinal double-spin asymmetry, $A_{LL}$, is sensitive to
the gluon polarization distribution $\Delta g(\xBj)$
\cite{ALL_theory_paper-1, *ALL_theory_paper0, *ALL_theory_paper}.
While $\Delta g(\xBj)$ in the range $0.05 < x < 0.2$ has become
more constrained \cite{DSSV, deltaGfit}, less is known for $\xBj < 0.05$.
As two protons are involved in the collision, there are two $x$ values.
We denote the larger $x$ value as $x_1$ and the smaller as $x_2$.
In quark-gluon scattering, $x_1$ is most often associated with the quark
and $x_2$ with the gluon, since gluons dominate proton distribution
functions at lower $x$. The production of $\pi^0$-mesons with
$0.8 < \eta < 2.0$ at $\sqrt{s}=200$ GeV covers approximately the range
$0.1 < x_1 < 0.5$ and $0.01 < x_2 < 0.33$, with $x_1$ and $x_2$
increasing with $p_T$.  Figure \ref{fig:x1x2} shows Bjorken $x_1$ and $x_2$
distributions for two representative $p_T$ bins, based on simulations using
PYTHIA 6.423 \cite{Pythia} with tune ``Pro-pT0'' \cite{Pythia2} utilizing the
CTEQ5L set of unpolarized parton distribution functions \cite{CTEQ5L}.

\begin{figure*}
\begin{center}
\includegraphics[width=0.95\textwidth]{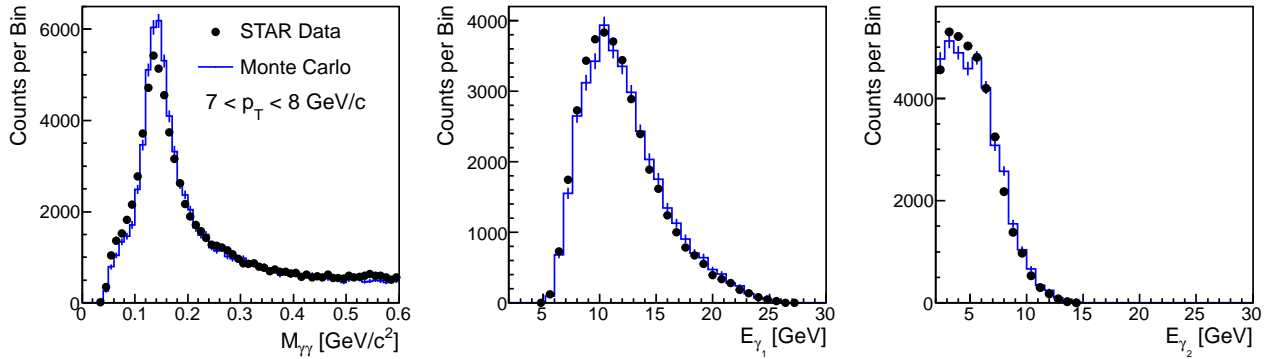}
\end{center}
\caption{\label{fig:dataMC}(Color online) Comparison of data to Monte Carlo
for the distributions of two-photon invariant mass (left) and energy for the
higher (center) and lower (right) energy photon. Distributions are shown with
a reconstructed transverse momentum range of $7 < p_T < 8$ GeV$/c$. For
the photon-energy distributions, a two-photon mass requirement of
$0.1<\Mgg<0.2$ GeV$/c^{2}$ is applied. The Monte Carlo distributions have
been normalized to the number of counts in the data distributions.}
\end{figure*}

Measurements of transverse single-spin asymmetries, $A_{N}$, for
inclusive $\pi^{0}$-production \cite{PHENIX_mid_AN}, as well as inclusive
jet production \cite{STAR_jet_AN2}, at central pseudorapidity have shown
no sizable effects. However, sizable asymmetries are observed for
inclusive $\pi^{0}$-production at forward pseudorapidity ($\eta\approx3$)
by several experiments over a wide range of $\sqrt{s}$ with no sign of
$\sqrt{s}$-dependence \cite{E704-88,*E704,STAR_FPD_AN1,
STAR_FPD2_xSec,STAR_FPD_AN2,FPD_lowPt,*FMS_xF,*FMS_xFpT,
*FMS_500,*FMS_200,PHENIX_AN62,*PHENIX_AN200pT,
*PHENIX_AN200pTxF}. It is expected that the Sivers and Collins effects
at twist-2 \cite{AN_Siv, *AN_Siv2, AN_Col, AN_Siv_Col,D'Alesio_Sivers3,
Yuan_Collins} as well as higher twist effects \cite{Efremov_twist3_1,
*Efremov_twist3_2,*Efremov_twist3_3, QiuSterman_twist3, Kanazawa10,
Kanazawa} contribute to these asymmetries (in particular at higher $\sqrt{s}$),
and measurements which map the dependence in $x_{F}$ and $p_{T}$ may
help elucidate the underlying mechanisms. While at large $p_T$, $A_N$ is
expected to scale as $1/p_T$ \cite{KPR,Yuan_twist3_1,Yuan_Collins,
D'Alesio_Sivers3}, previous results at forward pseudorapidity do not exhibit
this behavior \cite{STAR_FPD_AN2,FPD_lowPt,*FMS_xF,*FMS_xFpT,
*FMS_500,*FMS_200,PHENIX_AN62,*PHENIX_AN200pT,
*PHENIX_AN200pTxF,*PHENIX_AN200pTxF}. At intermediate $p_T$, the
behavior is unknown. Model predictions also differ on the expected
behavior of  $A_N$ as a function of $p_{T}$. For example, while a recent
model prediction based on the Collins effect in the color-glass condensate
formalism \cite{Yuan_Collins} expects a $1/p_{T}$ scaling modified by the
transverse-momentum dependence of the fragmentation function in the
unpolarized cross section, a recent twist-3 model \cite{Kanazawa} predicts
$A_{N}$ of a few percent at forward pseudorapidity that should persist out to
$p_{T}\sim15$ GeV$/c$.  The $A_N$ measurements described in this
paper cover the previously unmeasured region $0.06 < x_F < 0.27$ and
$5 < p_T < 12$ \GeVc.

\section{Analysis}

The data used for these measurements were taken with the STAR
detector \cite{STAR_NIM} during the 2006 RHIC run. The data for the
cross section were extracted from a sampled luminosity of 8.0 pb${}^{-1}$,
while the data for the longitudinal and transverse asymmetries were extracted
from sampled luminosities of 4.8 pb${}^{-1}$ and 2.8 pb${}^{-1}$, respectively.
The vertex positions were determined using charged particle tracks in
the time projection chamber (TPC) \cite{TPC_NIM}.  The beam-beam
counters (BBCs) \cite{BBC_paper2, *BBC_paper3} were used to
determine luminosity and were required in the event trigger.

The endcap electromagnetic calorimeter (EEMC) is used to measure the
energy  and position of photons from $\pi^0$ decays across the range of
$1.086<\eta_{\mathrm{det}}<2.00$, where $\eta_{\mathrm{det}}$ is the
detector $\eta$, relative to the nominal interaction point. The EEMC is a
lead-scintillator sampling calorimeter \cite{EEMC_NIM}, with both of the
first two layers and last layer being read out independently as preshower
and postshower layers, respectively. Each layer in the EEMC consists of
720 independent segments formed from 12 sections in pseudorapidity
($\eta$) and 60 sections in azimuth ($\phi$).  The segments in all layers
corresponding to a specific ($\eta$,$\phi$) range, when taken together,
are called a ``tower''.  A shower maximum detector (SMD) is located
between layers five and six (at a depth of $\sim5$ radiation lengths),
and consists of two layers of tightly packed triangularly shaped
scintillating strips $\sim1$ cm wide at the base.

Photons are reconstructed by first clustering the energy depositions
in the SMD strips to determine the position in $\eta$ and $\phi$, and then
using the corresponding EEMC towers to measure the photon energy.
The EEMC detector components are calibrated using the most probable
value of the Landau-peak response for minimum ionizing particles. Only
SMD energy clusters with at least 3 MeV of deposited energy and at least 2
MeV deposited in the central strip of the cluster, were used for this analysis.
Clusters are seven strips in size and are required to have at least five strips
with non-zero energy. The photon energy is determined by summing the
energy in a $3\times3$ set of towers. In the case where a given tower is
associated with more than one photon, the energy of the shared tower is
distributed between the photons in a manner proportional to the energy
each photon deposited in the SMD. Photons are further required to have
an energy of at least 2.0 GeV as measured in the associated tower(s) and
to be within the fiducial volume of $1.11 < \eta_{\mathrm{det}} < 1.96$.
The physical $\eta$, determined relative to the TPC-reconstructed primary
vertex, is required to be $0.8 < \eta < 2.0$. Further event selection
requirements are: (a) a valid bunch crossing (i.e. a bunch in both beams), (b)
a TPC-reconstructed vertex within $\pm 120$ cm of the nominal interaction
point, (c) a $\pi^0$ candidate transverse momentum $p_T > 5$ \GeVc, and
(d) a summed preshower energy for each photon tower cluster of less than
40 MeV to exclude spurious events, e.g., beam gas and other non-collision
backgrounds events. All possible pairs of photons that satisfy these
requirements are considered as $\pi^0$ candidates.

The invariant mass of photon pairs can be expressed as
\begin{equation}
 \label{eq:mgg}
	M_{\gamma\gamma}=\left(E_{\gamma_{1}}+E_{\gamma_{2}}\right)
	\sqrt{1-z^{2}_{\gamma\gamma}}\sin\frac{\theta_{\gamma\gamma}}{2},
\end{equation}
where $E_{\gamma_{1}}$ and  $E_{\gamma_{2}}$ represent the energies
of the two photons, $z_{\gamma\gamma}$ represents the two-photon
energy asymmetry $z_{\gamma\gamma}=\left|E_{\gamma_{1}}
-E_{\gamma_{2}}\right|/\left(E_{\gamma_{1}}+E_{\gamma_{2}}\right)$,
and $\theta_{\gamma\gamma}$ represents the opening angle between the
two photons. The limited photo-statistics in each SMD strip can cause a
cluster of energy deposited by a single shower to appear as two
clusters of energy and, thus, be reconstructed as two photons.  This
``false splitting'' effect accounts for a large fraction of $\pi^{0}$
candidates with invariant mass below 0.1 GeV/$c^{2}$.  False splitting
can be somewhat mitigated by a ``merging'' procedure. Simulation
studies indicate that when a false split results in multiple reconstructed
pion candidates with $p_{T} > 4$ \GeVc, the vast majority of candidates
are reconstructed within a radius $\sqrt{\Delta \eta^2 + \Delta \phi^2} < 0.05$.
Thus, if two $\pi^0$ candidates are found within a radius of $0.05$ then
these candidates are replaced with a new, merged candidate.
The momentum of the merged candidate is set to the sum of the momenta
of the contributing photons, without double counting photons that were
included in the original $\pi^0$ candidates. Simulations indicate a potential
loss of $\approx0.13\%$ of events with $p_{T} > 4$ GeV$/c$ from
merging two real pions, an effect considered negligible. The other
large contributor to low mass $\pi^0$ candidates is the case in which
one of the SMD clusters of a real $\pi^{0}$ is not reconstructed; and, thus,
the reconstructed photon from the real pion is never paired with the correct
second photon. The cluster may have been lost due to being
below the energy threshold or, more frequently, due to two clusters merging
in one of the layers. The real $\pi^0$ with the lost cluster will have its
opening angle, and thus its mass, reconstructed lower than the true value.

Reconstruction of $\pi^{0}$ candidates with invariant mass above
$0.2$ GeV$/c^{2}$ can arise from a conspiracy of two effects. Finite energy
resolution affects the reconstruction of $z_{\gamma\gamma}$. Furthermore,
when additional energy from the parent jet is deposited in the vicinity of the
photon pair, the reconstruction algorithm may include this energy with that
of the true pion. These two effects conspire to increase the amount of
$\pi^{0}$ signal reconstructed with mass above the peak region.

All events considered in this analysis are from a single trigger that
includes a coincidence requirement in the two BBCs, implying a $p+p$
collision.  The trigger requires at least one EEMC tower with transverse
energy above a given threshold and with the total transverse energy in
the $3 \times 3$ ``patch'' of towers surrounding and including the
high energy tower to be above a second threshold.  Although hardware
thresholds varied over the course of the data taking, the analysis
included an emulated trigger requirement, with thresholds of 4.3 GeV
and 6.2 GeV, respectively, for the high energy tower and the $3 \times 3$
tower patch.  These emulated trigger thresholds were 10\% above the
maximum hardware triggers. $\pi^{0}$ candidates with $p_{T}$
below the software energy threshold can arise from several sources,
e.g., the spread and offset from the nominal longitudinal position
of the collision vertex, off-line rejection from the $\pi^{0}$ candidate of
hadronic energy deposits, and events with $\pi^{0}$ candidates not
associated with the tower or tower clusters firing the trigger.

\begin{figure}
\begin{center}
\includegraphics[width=0.47\textwidth]
{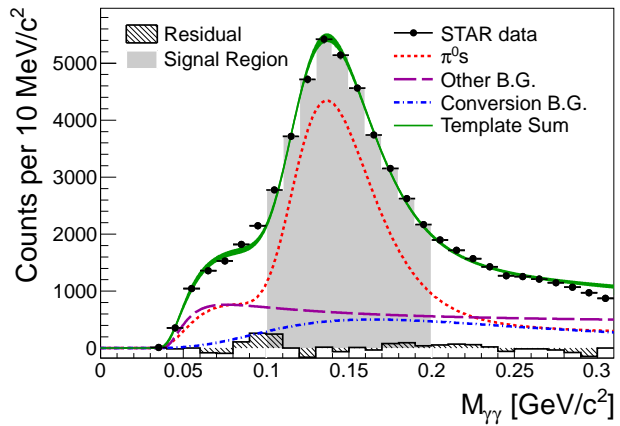}
\end{center}
\caption{\label{fig:mass}(Color online) Invariant mass distribution for the
two-photon system with $7 < p_T < 8$ \GeVc. Also included on the
plot are the template functions for the signal and two backgrounds
(scaled and shifted according to the fit results), the residual
between the data and the sum of the templates, and a gray-shaded area
indicating the peak region. }
\end{figure}

To understand the effects of backgrounds, efficiencies, and $p_{T}$ resolution,
data have been compared to a Monte Carlo simulation based on PYTHIA, as
described previously, with GEANT 3.21 \cite{GEANT0,*GEANT1} to model
detector response. An example of the data-Monte Carlo studies is shown in
Fig. \ref{fig:dataMC}. In this example, distributions are compared between
two-photon invariant mass and single-photon energy for two-photon events
with a reconstructed transverse momentum range of $7 < p_{T} < 8$ \GeVc.
In general, data and Monte Carlo distributions show reasonable agreement
for $p_{T} > 6$ \GeVc. For $p_{T} < 6$ \GeVc, discrepancies between data
and Monte Carlo lead to increased, but well-constrained systematic
uncertainties in the estimation of signal fractions.

\begin{figure*}
\begin{center}
\includegraphics[width=0.95\textwidth]{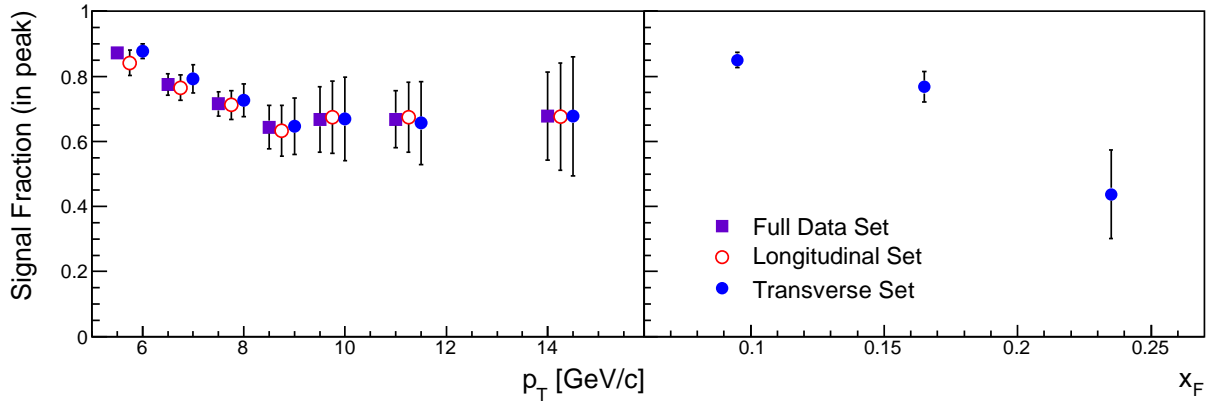}
\end{center}
\caption{\label{fig:sigFrac}(Color online) Signal fractions calculated within
the ``peak region'' of $0.1<\Mgg<0.2$ \GeVcc. Fractions for the full dataset
as well as the subsets of longitudinal and transverse polarizations are
shown as a function of $p_{T}$ (left), and the fractions for the transversely
polarized data are shown as a function of $x_{F}$ integrated over
$5 < p_{T} < 12$ GeV$/c$ (right). Uncertainties on the signal fractions
arise from those of the template forms determined from Monte Carlo and
from their application to the data. The size of the uncertainties is influenced
by the number of events in the available data and Monte Carlo and the
quality of fits to Monte Carlo and data. The same Monte Carlo sample is
used to extract the signal fractions for the three datasets.}
\end{figure*}

The signal fraction was determined by fitting a linear combination of
template functions to the two-photon invariant mass distribution over the
range $0 < \Mgg < 0.3$ GeV$/{}c^2$ for each $p_T$ (or $x_F$) bin. Three
template functions were determined by fitting the functions to Monte Carlo
data to represent (a) the $\pi^0$ signal, (b) the conversion background
where the two reconstructed ``photons'' that formed the $\pi^0$ candidate
were actually the two leptons from a photon that converted in material
upstream of the EEMC, and (c) all other backgrounds, including
combinatoric backgrounds. Signal and conversion background events were
determined by matching the momentum direction of reconstructed pairs to
that of generated $\pi^{0}$'s and decay photons, respectively, in
$\left(\eta,\phi\right)$ space. Non-matched reconstructed pairs were
considered ``other'' backgrounds. The shapes of the template functions
were chosen to match the shapes of the  various contributions from Monte
Carlo. For the $\pi^{0}$ signal the sum of two skewed Gaussian distributions
was chosen, while the two background contributions were each represented
by single skewed Gaussian distributions. The parameter values were fixed
by fitting the template functions to the contributions in Monte Carlo, and the
relative weights of the templates were determined by fitting a linear
combination of the template functions to the data. When fitting the weights of
the three template functions an additional factor was also included to account
for the energy scale difference between the data and the Monte Carlo. This
energy scale difference was not simply related to the calibration, but was also
affected by assumptions about the sampling fraction used in the simulation.
The energy scale difference extracted from the fits is approximately $3\%$.

The data and template functions for the $7 < p_T < 8$ GeV$/c$ bin are
shown in Fig. \ref{fig:mass}.  While the fits to determine the signal
fraction cover $0 < \Mgg < 0.3$ GeV$/c^{2}$, only $\pi^0$ candidates
with $\Mgg$ in the range $0.1 < \Mgg < 0.2$ GeV$/c^{2}$ (defined as
the peak region) were used for the remainder of the analysis. The signal
fraction in the peak region (Fig. \ref{fig:sigFrac}) was computed from the
weights, the data versus simulation energy scale factor, and integrals of
the template functions.  The product of the signal fraction in the peak region
and the number of $\pi^0$ counts within this region then gives the number
of background-subtracted $\pi^0$'s for the given bin.

To compute the cross section, the number of background-subtracted
$\pi^0$'s was corrected for $p_T$ bin smearing by applying the inverse
of a smearing matrix, obtained from the same PYTHIA Monte Carlo data
set as used above.  The final cross section was then computed using
\begin{equation}\
  \label{eq:crossSection}
  E \frac{d^{3}\sigma}{d\bm p^3} = \frac{1}{\Delta \phi\ \Delta \eta\ \Delta p_T}
  \frac{1}{\langle p_T \rangle} \frac{1}{\textnormal{BR}} \frac{1}{\epsilon}
  \frac{N}{\mathcal L},
\end{equation}
where $N$ is the corrected number of $\pi^0$'s, $\mathcal{L}$ is the
sampled luminosity (including dead-time corrections), $\epsilon$ is
the product of reconstruction and trigger efficiencies, BR is the
branching ratio $\pi^0\rightarrow \gamma\gamma$ \cite{PDG}, $\langle
p_T \rangle$ is the average $p_T$ for the particular $p_T$ bin,
$\Delta p_T$ is the width of the $p_T$ bin, and $\Delta \phi$ (equal
to $2\pi$) and $\Delta \eta$ (equal to 1.2) are the $\phi$ and $\eta$
phase space factors.  The trigger efficiency is below 10\% for
$\pi^0$'s with $5 < p_T < 6$ \GeVc, and plateaus above 40\% at $p_T
\approx 9$~GeV/$c$.  The reconstruction efficiency is around 30\% for
$5<p_T<9$~GeV/$c$, and decreases to around 20\% for $12 < p_T <
16$~GeV/$c$.

\begin{figure}
\begin{center}
\includegraphics[width=0.49\textwidth]
{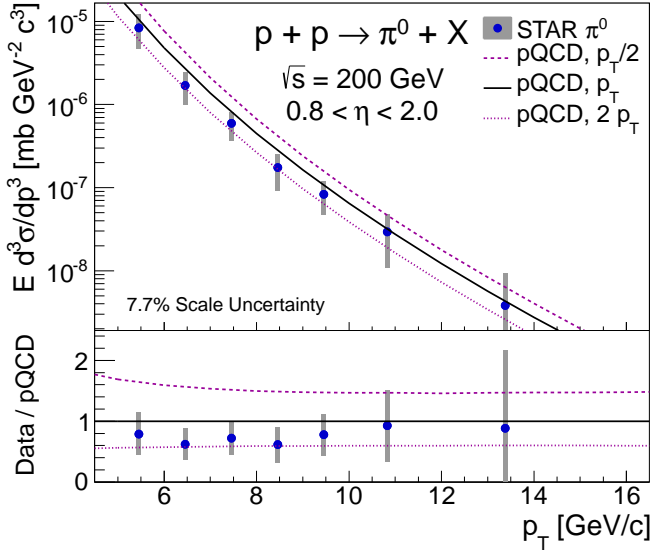}
\end{center}
\caption{\label{fig:xSec2}(Color online) Upper panel: the $\pi^0$ cross
section (blue markers) is shown compared with an NLO pQCD calculation
\cite{NLOcorrectionsPion} with three options for the scale parameter.
Statistical uncertainties are shown by the error bars which are
indistinguishable from the markers in all bins.  Systematic uncertainties
are shown by the error boxes.  The lower panel presents the ratio of the
data to the $p_T$-scale theory curve, as well as the ratio of the
$2p_T$-scale and $p_T/2$-scale theory curves to the $p_T$-scale curve. }
\end{figure}

The longitudinal spin asymmetries were computed by subtracting the
luminosity asymmetry from the asymmetry in the number of $\pi^0$
candidates and dividing this difference by the luminosity-weighted
polarization.  Specifically, one can write
\begin{eqnarray}
\label{eq:ALL}
  A_{LL} &=& \frac{1}{\left\langle P_B P_Y \right\rangle}
  \bigg(\frac{N^{++} - N^{+-} - N^{-+} + N^{--}}
  {N^{++} + N^{+-} + N^{-+} + N^{--}} \nonumber\\&&
   {}-\frac{L^{++} - L^{+-} - L^{-+} + L^{--}}
   {L^{++} + L^{+-} + L^{-+} + L^{--}}\bigg),
\end{eqnarray}
\begin{eqnarray}
\label{eq:ALB}
  A_{L,B} &=& \frac{1}{\left\langle P_B \right\rangle}
  \bigg(\frac{N^{++} + N^{+-} - N^{-+} - N^{--}}
  {N^{++} + N^{+-} + N^{-+} + N^{--}} \nonumber\\&&
  {}-\frac{L^{++} + L^{+-} - L^{-+} - L^{--}}
  {L^{++} + L^{+-} + L^{-+} + L^{--}}\bigg), \\
\label{eq:ALY}
  A_{L,Y} &=& \frac{1}{\left\langle P_Y \right\rangle}
  \bigg(\frac{N^{++} - N^{+-} + N^{-+} - N^{--}}
  {N^{++} + N^{+-} + N^{-+} + N^{--}} \nonumber\\&&
  {}-\frac{L^{++} - L^{+-} + L^{-+} - L^{--}}
  {L^{++} + L^{+-} + L^{-+} + L^{--}}\bigg).
\end{eqnarray}
Here, subscripts $B$ and $Y$ represent the blue (momentum from
the interaction region towards the EEMC) and yellow (momentum
aimed away from the EEMC) beams, $N$ denotes the number of
counts in the signal region, and $L$ indicates the luminosity. The
superscripts $+$ and $-$ designate the longitudinal polarization
directions of the blue beam and yellow beams, respectively.
Equations \ref{eq:ALB}, \ref{eq:ALY}, and \ref{eq:ALL} assume
negligible contributions from terms of the form
\begin{eqnarray}
A_{L,B}\times\frac{L^{++}-L^{--}-L^{+-}+L^{-+}}{L^{++}+L^{--}+L^{+-}+L^{-+}}
\end{eqnarray}
(similarly for $A_{L,Y}$) and also from terms coupling $A_{LL}$ to the
luminosity asymmetry. Luminosity asymmetries are kept quite small
due to the ability of RHIC to alternate spin directions for successive
bunch patterns using a complex 8-bunch polarization pattern. Since
the parity-violating asymmetry $A_{L}$ is expected to be quite small,
these correction terms are considered negligible. The spin-dependent
luminosities are calculated from the sum of BBC coincidences over a
run, after sorting bunches for each spin combination. The
luminosity-weighted average polarizations for the longitudinally
polarized data have values
$\left\langle P_B \right\rangle = 0.56$ and
$\left\langle P_Y \right\rangle = 0.59$,
and the luminosity-weighted average product of the polarizations has the
value $\left\langle P_B P_Y \right\rangle = 0.33$. The relative polarization
uncertainty of each beam is $4\%$, and the relative uncertainty for the
product is $6\%$.

\begin{figure}
\begin{center}
\includegraphics[width=0.49\textwidth]
{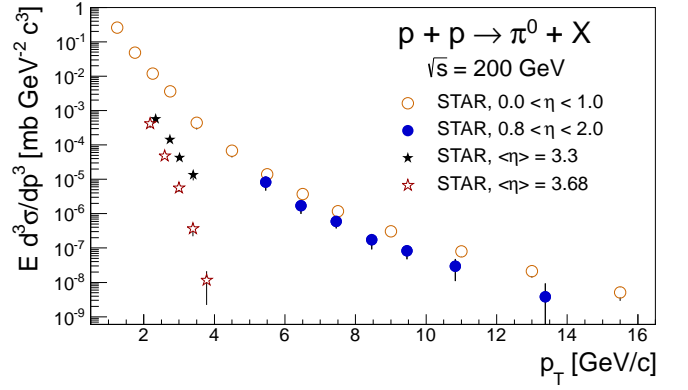}
\end{center}
\caption{\label{fig:xSec1}(Color online) The $\pi^0$ cross section at
various ranges of pseudorapidity as measured by STAR.  Error bars
indicate the total uncertainty. The closed blue circles are the results of
this analysis, while the other points are previously published results that
use the STAR barrel electromagnetic calorimeter (open orange circles)
\cite{STAR_pi0_BEMC} and the forward pion detectors (closed black
stars and open red stars) \cite{STAR_FPD1_xSec,STAR_FPD2_xSec}.}
\end{figure}

The signal fraction was determined using data summed over the spin
states.  The asymmetries were corrected for the background asymmetry
using
\begin{equation}
  \label{eq:asymBkgSub}
  A^{sig} = \frac{1}{s}\left( A^{raw} - (1-s) A^{bkg} \right),
\end{equation}
where $s$ is the signal fraction, $A^{sig}$ is the asymmetry of the
$\pi^0$ signal, $A^{raw}$ is the asymmetry value before background
subtraction (Eqs. \ref{eq:ALL}, \ref{eq:ALB}, and \ref{eq:ALY}), and
$A^{bkg}$ is an estimate of the background asymmetry.  The background
asymmetries were estimated as the average of the $p_{T}$-integrated
asymmetries in two sideband regions ($0<\Mgg<0.1$ GeV$/c^{2}$ and
$0.2<\Mgg<0.3$ \GeVcc), and were found to be less than $1\sigma$
from zero, with $\sigma \approx 0.01$.

The transverse spin asymmetry was computed by binning with respect to
$\phi$, the angle between the azimuthal angles of the $\pi^0$ and the
spin polarization vector.  The raw cross ratio $\mathcal E(\phi)$ was
computed per $\phi$ bin,
\begin{eqnarray}
  \label{eq:mathcalAN1}
  \mathcal E(\phi) &=&
  \frac{\sqrt{N^{\uparrow}\left(\phi\right)N^{\downarrow}\left(\phi+\pi\right)}
  -\sqrt{N^{\downarrow}\left(\phi\right)N^{\uparrow}\left(\phi+\pi\right)}}
  {\sqrt{N^{\uparrow}\left(\phi\right)N^{\downarrow}\left(\phi+\pi\right)}
  +\sqrt{N^{\downarrow}\left(\phi\right)N^{\uparrow}\left(\phi+\pi\right)}},
  \nonumber\\
\end{eqnarray}
where $N$ represents the number of counts, $\uparrow$ denotes beam
spin polarized vertically upward in the lab frame, and $\downarrow$
denotes beam spin polarized vertically downward in the lab frame.
The quantity $\mathcal E(\phi)$ was fit to the equation
$C+\varepsilon\sin\phi$, the background was subtracted using
Eq. \ref{eq:asymBkgSub} with $A^{raw}=\varepsilon$, and the final
result for $A_N$ was obtained by dividing by the luminosity weighted
polarization.  The luminosity-weighted average polarizations for the
transversely polarized data have values
$\left\langle P_B \right\rangle = 0.54$ and
$\left\langle P_Y \right\rangle = 0.55$. The uncertainty due to
propagation of the relative polarization uncertainty of each beam is
$4\%$ \cite{CNIPol0,*CNIPol1}. The background asymmetries were
estimated as the average of the asymmetry in the two sideband regions,
and were found for both $A_{N}$ and $A_{LL}$ to be less than
$1\sigma$ from zero, again with $\sigma \approx 0.01$.

\begin{figure}
\begin{center}
\includegraphics[width=0.49\textwidth]
{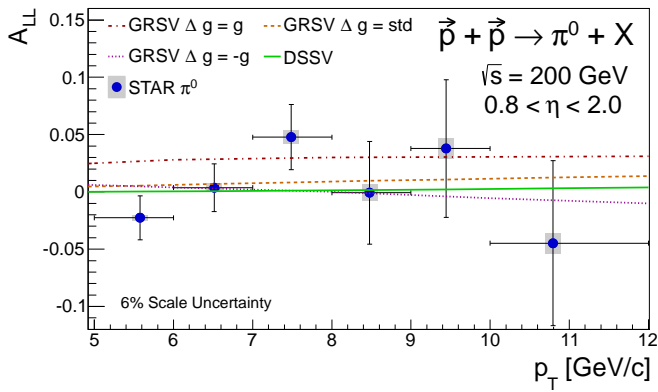}
\end{center}
\caption{\label{fig:ALL}(Color online) The $A_{LL}$ results (blue
markers) are presented with the DSSV prediction \cite{DSSV} and the
GRSV prediction \cite{GRSV} using the best fit to polarized DIS
($\Delta g=\mathrm{std}$) and the maximum and minimum allowed
values for gluon polarization. Statistical uncertainties are shown by the
error bars, whereas systematic uncertainties are indicated by the error
boxes. The $6\%$ scale uncertainty is due to beam polarization
uncertainty.}
\end{figure}

\section{Results}

\subsection{Cross Section}

Figure \ref{fig:xSec2} presents the measured cross section for neutral
pions produced over the transverse momentum range $5 < p_{T} < 16$
GeV$/c$. Contributions to the systematic uncertainties include those
related to the uncertainty on the signal fraction, the smearing matrix, the
effect of repeating the analysis with an additional $4 < p_T < 5$ GeV/$c$
bin, the reconstruction and trigger efficiencies, the EEMC energy resolution,
and the overall EEMC energy scale. The signal fraction uncertainty
includes contributions from the uncertainties on the parameters in the
template functions, the uncertainty on the weights of the templates, the
uncertainty on the scale parameter and its effect on the integrals used
to determine the signal fraction in the peak, and a contribution based
on the integral of the residual in the signal region. Uncertainty on the
luminosity results in a $7.7\%$ vertical scale uncertainty. The dominant
uncertainty on the cross section is the overall energy scale uncertainty,
which is correlated over all bins.

The measured cross section results in Fig. \ref{fig:xSec2} are compared to
a theory prediction based on NLO pQCD and global fits of distribution and
fragmentation functions \cite{NLOcorrectionsPion}. The CTEQ6.5 set of
parton distribution functions \cite{CTEQ65} and DSS fragmentation functions
\cite{DSS} are used. The EEMC $\pi^0$ cross section data points are
observed to lie between the calculations that set the factorization,
renormalization, and fragmentation scales to $p_T$ and $2p_T$. This is
qualitatively consistent with central pseudorapidity measurements from
PHENIX, both in published results at $\sqrt{s}=200$ GeV
\cite{PHENIX_pi0_xSec200} and preliminary results at $\sqrt{s}=500$ GeV
\cite{PHENIX_pi0_xSec500}. In each of these measurements, the cross
section is lower than the $p_T$-scale theory curve in the region of
$5 < p_T < 16$ \GeVc. Within uncertainties, previous STAR results at
$\sqrt{s}=200$ GeV are in good agreement with the $p_{T}$-scale theory
predictions \cite{STAR_pi0_BEMC}.

Figure \ref{fig:xSec1} shows the cross section results of this
analysis in comparison with previously published STAR results in other
pseudorapidity and transverse-momentum regions. While the entire STAR
detector has a broad range of coverage, the results presented here lie in a
previously unmeasured region. The results indicate that the cross
section changes slowly with respect to $\eta$ at lower $\eta$ and has
significant $\eta$ dependence at higher $\eta$, with the transition
lying between $\eta = 2$ and $\eta = 3.3$.

\begin{figure}
\begin{center}
\includegraphics[width=0.49\textwidth]{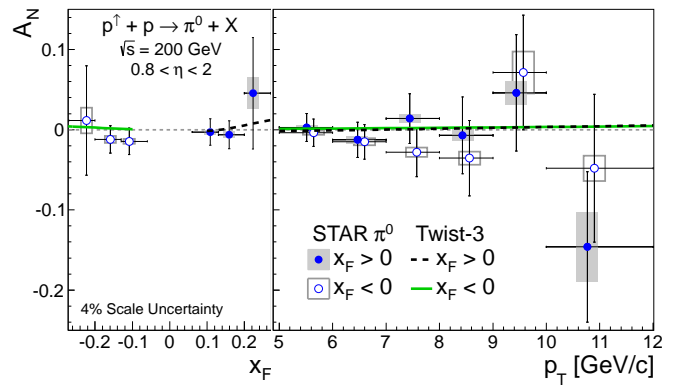}
\end{center}
\caption{\label{fig:AN1}(Color online) The $A_N$ results are plotted versus
$x_F$ integrated over $5 < p_{T} < 12$ GeV$/c$ (left panel) and versus
$p_T$ integrated over $0.06 < \left|x_{F}\right| < 0.27$ (right panel).
Statistical uncertainties are shown by error bars, whereas systematic
uncertainties are indicated by error boxes.  Negative $x_F$ results are
depicted with open circles and open error boxes, while positive $x_F$ results
are exhibited with closed circles and closed systematic error boxes. The
$A_{N}$ results are presented with model predictions based on the twist-3
mechanism in the collinear factorization framework \cite{Kanazawa}.
The $4\%$ scale uncertainty is due to beam polarization uncertainty.}
\end{figure}

\subsection{Longitudinal Asymmetries}

The $A_{LL}$ results for $5 < p_T < 12$ GeV$/c$ are shown in
Fig. \ref{fig:ALL}.  Systematic uncertainties include those on the signal
fraction and on the estimate of the background asymmetry. The relative
luminosity uncertainty was found to be negligible compared to the
systematic uncertainties from the signal fraction and the background
asymmetry. Integrating over $5 < p_T < 12$ GeV$/c$ yields a value of
$A_{LL} = 0.002 \pm 0.012$. Uncertainty in the product of beam
polarizations results in a $6\%$ vertical scale uncertainty as indicated
in the figure. This systematic uncertainty is correlated across all bins
and vanishes as the measured asymmetries go to zero.

Model predictions, based on global fits by the GRSV group to polarized
deep inelastic scattering (DIS) data \cite{GRSV} and global fits by the DSSV
group to polarized DIS, semi-inclusive DIS, and proton-proton collisions
\cite{DSSV}, are shown along with the measured $A_{LL}$ results in
Fig. \ref{fig:ALL}. For the GRSV prediction, calculations are shown for the
best fit to polarized DIS ($\Delta g=\mathrm{std}$) as well as those for the
maximum ($\Delta g=g$) and minimum ($\Delta g=-g$) allowed gluon
polarization. Both GRSV and DSSV are calculated at NLO. DSS
fragmentation functions \cite{DSS} are utilized, as well as the CTEQ6.5 set
of parton distribution functions \cite{CTEQ65} with the unpolarized
NLO calculation \cite{NLOcorrectionsPion}. The $A_{LL}$ results lack
the precision to distinguish between the present various parameterizations
of gluon polarization, yet, may still impact global extractions of
$\Delta g\left(x\right)$ which reach to less-constrained values of low
Bjorken-$x$ or those not presently including RHIC data
(e.g. Ref. \cite{NNPDF}).

The parity-violating single-spin asymmetry, $A_{L}$, was also measured
for each of the colliding beams and is consistent with zero. Integrating
over $p_T$ from $5 < p_T < 12$ GeV$/c$ yields $A_{L} = -0.003 \pm 0.007$
(blue beam) and $A_{L} = -0.001 \pm 0.007$ (yellow beam).

\begin{figure}
\begin{center}
\includegraphics[width=0.48\textwidth]{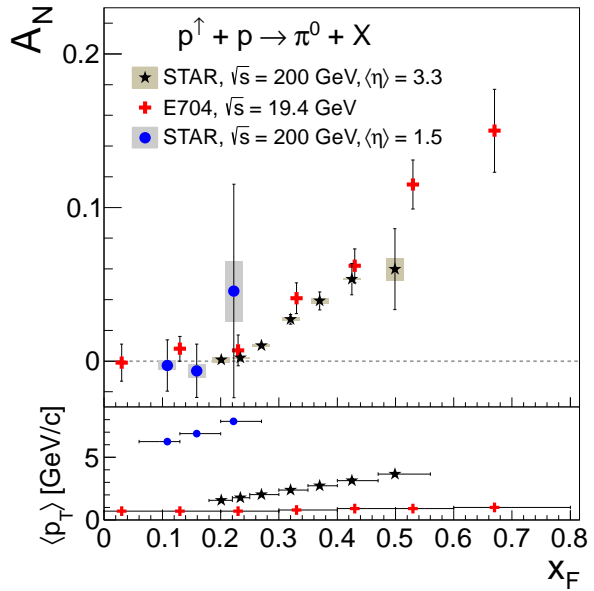}
\end{center}
\caption{\label{fig:AN2}(Color online) The present $A_N$ results (blue
circles) are compared with previously published values of $A_N$
\cite{E704-88,*E704,STAR_FPD_AN2} as a function of $x_F$ (top panel).
The average $p_T$ values within each $x_F$ bin are compared for the
various measurements (bottom panel).}
\end{figure}

\subsection{Transverse Spin Asymmetries}

The results for $A_N$ versus $x_F$, integrated over $5 < p_T < 12$
GeV/$c$, as well as $A_N$ versus $p_T$, integrated over
$0.06 < \left|x_{F}\right| < 0.27$, are shown in Fig.  \ref{fig:AN1}. Asymmetries
for $x_{F}>0$ are measured accounting for the polarization direction of the
blue beam, while those for $x_{F}<0$ are measured accounting for that of the
yellow beam. Systematic uncertainties include those on the signal fraction, on
the estimate of the background asymmetry, and on single-beam backgrounds.
Uncertainty in the beam polarizations results in a $4\%$ vertical scale
uncertainty as indicated in the figure. Over the $x_F$ region of this measurement, $A_N$ is
statistically consistent with zero and no strong conclusions about the $p_T$
dependence can be made. The measured asymmetries are presented with
model predictions based on the twist-3 mechanism in the collinear
factorization scheme \cite{Kanazawa}. The measured asymmetries are
consistent with the model predictions which expect small effects for both
$x_{F} > 0$ and $x_{F} < 0$.

The present $A_N$ results are compared with previously published results
in Fig.~\ref{fig:AN2}. The lower panel of Fig.~\ref{fig:AN2} shows the average
$p_{T}$ for each bin of $x_{F}$. As anticipated from the previous results at
lower $p_T$ and similar $x_F$ \cite{E704-88,*E704,STAR_FPD_AN1,
STAR_FPD_AN2,FPD_lowPt,PHENIX_AN62}, $A_N$ is statistically consistent
with zero. Integrating over $0.06 < \left|x_{F}\right| < 0.27$ over the
aforementioned range of $p_{T}$ yields $A_N = 0.000 \pm 0.009$
for $x_F > 0$ and $A_N = 0.009 \pm 0.009$ for $x_F<0$, with
$\langle |x_F| \rangle = 0.14$.

\section{Conclusions}

Neutral pions produced from polarized proton-proton collisions with 
$\sqrt{s} = 200$ GeV at RHIC have been detected using the STAR
endcap electromagnetic calorimeter. The production cross section,
the longitudinal double and single-spin asymmetries, and the
transverse single-spin asymmetry have been measured for $\pi^0$'s with
$0.8 < \eta < 2.0$. The spin asymmetries were extracted for $\pi^{0}$'s over
the range $5 < p_T < 12$ GeV$/c$, while the cross section was measured for
those over the range $5 < p_T < 16$ GeV$/c$. These results probe a
region of phase space not previously studied at RHIC energies,
complementing measurements in neighboring regions. The cross
section is slightly lower than previously published measurements at
more central ranges of pseudorapidity and within the scale uncertainty
of a pQCD-calculated prediction. The $A_{LL}$ measurement is
compared with a model prediction and includes data with Bjorken
$x_2$ reaching below 0.01 based on calculations utilizing unpolarized
parton distribution functions. The measured values of the parity-violating
spin asymmetry, $A_L$, are consistent with zero. The measured values
of $A_N$ are compared with a twist-3 model prediction and found to be
consistent. The present results are
also compared with previously published measurements which also
suggest small asymmetries for similar $x_{F}$ and lower values of $p_{T}$.

\begin{acknowledgments}
The authors thank M. Stratmann, W. Vogelsang, and K. Kanazawa for
providing calculations and discussion. We thank the RHIC Operations
Group and RCF at BNL, the NERSC Center at LBNL, the KISTI Center
in Korea and the Open Science Grid consortium for providing resources
and support. This work was supported in part by the Offices of NP and
HEP within the U.S. DOE Office of Science, the U.S. NSF, CNRS/IN2P3,
FAPESP CNPq of Brazil, Ministry of Ed. and Sci. of the Russian Federation,
NNSFC, CAS, MoST and MoE of China, the Korean Research Foundation,
GA and MSMT of the Czech Republic, FIAS of Germany, DAE, DST, and
CSIR of India, National Science Centre of Poland, National Research
Foundation (NRF-2012004024), Ministry of Sci., Ed. and Sports of the
Rep. of Croatia, and RosAtom of Russia. Finally, we gratefully
acknowledge a sponsored research grant for the 2006 run period from
Renaissance Technologies Corporation.
\end{acknowledgments}

\bibliography{pi0-EEMC-2006}

\end{document}